\documentclass[fleqn,12pt]{article}
\usepackage[dvips]{graphicx}
\usepackage{cite}
\usepackage{times}
\renewcommand{\refname}{\normalsize References}

\makeatletter
\renewenvironment{thebibliography}[1]
{\section*{\refname\@mkboth{\refname}{\refname}}\vspace{4pt}%
   \list{\@biblabel{\@arabic\c@enumiv}}%
        {\settowidth\labelwidth{\@biblabel{#1}}%
         \leftmargin\labelwidth
         \advance\leftmargin\labelsep
	 \setlength\itemsep{0pt}
	 \setlength\baselineskip{14pt}
         \@openbib@code
         \usecounter{enumiv}%
         \let\p@enumiv\@empty
         \renewcommand\theenumiv{\@arabic\c@enumiv}}%
   \sloppy
   \clubpenalty4000
   \@clubpenalty\clubpenalty
   \widowpenalty4000%
   \sfcode`\.\@m}
  {\def\@noitemerr
    {\@latex@warning{Empty `thebibliography' environment}}%
   \endlist}
 
\makeatletter
\long\def\abstract#1{\def\@abstract{#1}}%
\def\@abstract{}
\long\def\affiliation#1{\def\@affiliation{#1}}%
\def\@affiliation{}
\long\def\email#1{\def\@email{#1}}%
\def\@email{}
\def\@maketitle{%
\begin{center}
\vspace*{1pt}
{\large\bfseries\@title}
\\[12pt]
\@author
\\[12pt]
{\textit{\@affiliation}}
\\
\@email
\end{center}
\leftmargini15mm
\begin{quotation}\noindent {\bfseries\abstractname}\\
\@abstract\end{quotation}%
 \vskip14pt}%
\makeatother

\makeatletter
\renewcommand{\section}{\@startsection
{section}{3}{0mm}{5mm}{0.1pt}{\bfseries \normalsize}}
\makeatother

\makeatletter
\renewcommand{\subsection}{\@startsection
{subsection}{3}{0mm}{5mm}{0.01pt}{\bfseries \normalsize}}
\makeatother

\title{%
Detection of the Direct Hyperfine Transition of Positronium Atoms using sub-THz High-power Radiation
}

\author{T.~Suehara$^1$, A.~Miyazaki$^2$, T.~Yamazaki$^2$, G.~Akimoto$^2$, A.~Ishida$^2$, T.~Namba$^1$, S.~Asai$^2$, T.~Kobayashi$^1$,
	H.~Saito$^3$, M.~Yoshida$^4$, T.~Idehara$^5$, I.~Ogawa$^5$, Y.~Urushizaki$^5$ and S.~Sabchevski$^6$}

\affiliation{%
$^1$ International Center for Elementary Particle Physics (ICEPP), The University of Tokyo, 7-3-1 Hongo, Bunkyo-ku, Tokyo, 113-0033, Japan \\
$^2$ Department of Physics, Graduate School of Science, The University of Tokyo, 7-3-1 Hongo, Bunkyo-ku, Tokyo, 133-0033, Japan \\
$^3$ Department of General Systems Studies, Graduate School of Arts and Sciences, The University of Tokyo, 3-8-1 Komaba, Meguro-ku, Tokyo, 153-8902, Japan \\
$^4$ Accelerator Laboratory, High Energy Accelerator Research Organization (KEK), 1-1 Oho, Tsukuba, Ibaraki, 305-0801, Japan \\
$^5$ Research Center for Development of Far-Infrared Region, University of Fukui (FIR-FU), 3-9-1 Bunkyo, Fukui, Fukui, 910-8507, Japan \\
$^6$ Institute of Electronics of the Bulgarian Academy of Science, 1784 Sofia, Bulgaria
}

\email{%
suehara@icepp.s.u-tokyo.ac.jp
}

\abstract{
Hyperfine splitting of positronium is an important parameter
for particle physics. 
This paper gives experimental techniques and results of R\&D studies
of our experiment to observe direct hyperfine transition of ortho-positronium
to para-positronium.
}

\begin{document}
\maketitle
\thispagestyle{empty}
\pagestyle{empty}

\section{Introduction}
Positronium is an ideal system for the research of the bound state quantum electrodynamics (QED). 
The energy difference between ortho-positronium (o-Ps, $1^3S_1$ state) and
para-positronium (p-Ps, $1^1S_0$ state), hyperfine splitting of positronium (Ps-HFS),
is a good measure for QED validation and also good for a search of unknown phenomena.
Previous experimental results of the Ps-HFS show
3.9$\sigma$ (15 ppm) discrepancy from the QED calculation.
In order to investigate this discrepancy,
we are preparing a new experiment to observe the direct transition of
ortho-positronium (o-Ps) to para-positronium (p-Ps) 
Theoretical and experimental context of HFS measurements are summarized in \cite{asai}.


Direct HFS transition is caused by applying photons whose energy is just at HFS (203 GHz).
We can obtain HFS value by measuring HFS transition rate at several frequencies
throughout the Breit-Wigner resonance curve (1.5 GHz FWHM) around the 203 GHz peak.
The method of the HFS transition detection is described in Section 3.

The most challenging issue for the direct HFS observation is the photon supply. 
Since Ps-HFS is `M1' transition, which is prohibited at non-relativistic limit,
the natural transition rate is only $3 \times 10^{-9} \mathrm{sec}^{-1}$, 
which results in very small cross section of the photon-induced HFS transition.
Therefore, extremely high photon density of $> 10^{15} \mathrm{cm}^{-3}$ is needed to observe the transition.
Furthermore, the light source should be frequency-tunable by several GHz to obtain the resonance curve.
We plan to use a sub-THz gyrotron with a Fabry-P\'erot cavity to meet the requirements.

\section{Optical Design}

\begin{figure}[tb]
\begin{center}
\includegraphics[width=150mm]{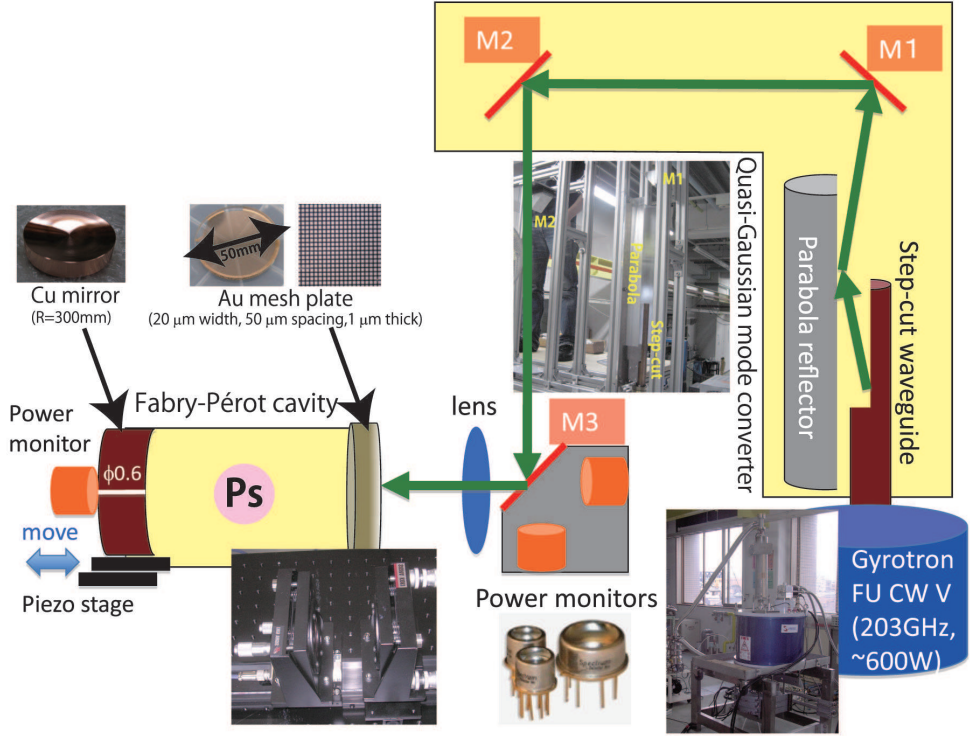}
\label{fig:optical-design}
\caption{Optical design of our experiment.}
\end{center}
\end{figure}

Fig.~\ref{fig:optical-design} shows an overall design of our setup to accumulate 203 GHz photons.
The setup consists of (1) a 203 GHz gyrotron, (2) quasi-Gaussian mode converter and lens, and
(3) Fabry-P\'erot cavity.

\subsection{Gyrotron}

We plan to use a gyrotron `Gyrotron FU CW V',
which is fabricated just for our HFS experiment.
The gyrotron has 203 GHz resonant frequency and 609 W output power (measured) at the window.
The output radiation mode is TE03, which is well-suited to be converted into quasi-Gaussian mode.
Since the gyrotron currently has no specific mechanism to change its resonant frequency,
tunable frequency is limited by the gyrotron cavity to several handreds of MHz.

\subsection{Quasi-Gaussian Mode Converter and Lens}

\begin{figure}[tb]
\begin{center}
\begin{minipage}{14pc}
\includegraphics[width=13pc]{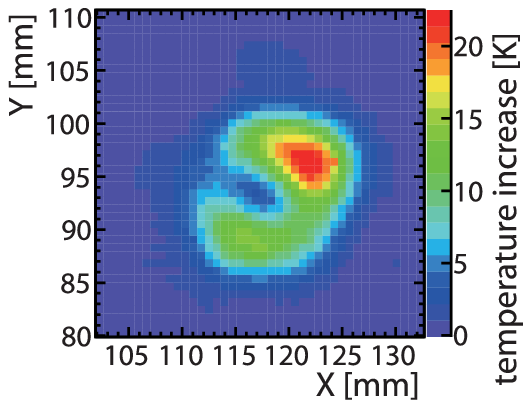}
\end{minipage}\hspace{2pc}%
\begin{minipage}{14pc}
\includegraphics[width=13pc]{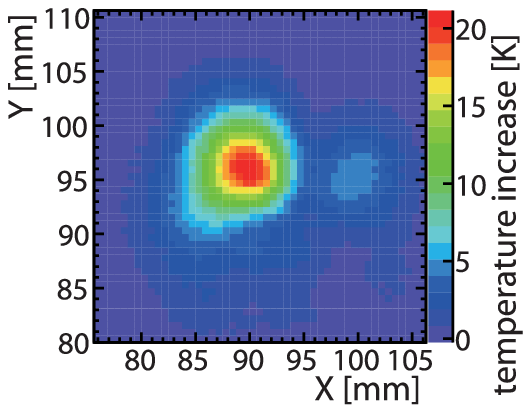}
\end{minipage} 
\caption{\label{fig:powerdist}Beam profile after the focusing lens without (left) and with (right) the Gaussian mode converter.}
\end{center}
\end{figure}

To introduce the gyrotron power to the Fabry-P\'erot cavity with high efficiency,
input transverse mode to the cavity must be TEM00-like.
Our mode converter consists of a step-cut waveguide, `Vlasov' parabola reflector
and two parabola mirrors. The design is derived from \cite{vlasov}.
Output of the `Vlasov' mirror should be bi-Gaussian at far field, which is converted
to Gaussian by two other parabola mirrors.
After the converter is located a lens, which controls size, dispersion angle and position
of the photon beam at the cavity entrance with 5-axis micro-mover,
in order to maximize fraction of the power to enter the cavity.
Fig.~\ref{fig:powerdist} shows obtained power distributions after the lens,
both with and without the Gaussian mode converter.
The power distribution with the mode converter is apparently more Gaussian-like.

\subsection{Fabry-P\'erot Cavity}

The Fabry-P\'erot cavity consists of a metal-mesh mirror as the input mirror
and a copper concave mirror at the other side. Cavity length is precisely controlled by a piezo-walk stage.
The mesh parameters are tuned
using electromagnetic field simulation to achieve $> 99\%$ round-trip reflection (already confirmed by measurements)
with minimum loss at power introduction.
Details of the optimization are described in \cite{posmol}.

For efficient introduction of the power, mode matching between input and inner field is critical.
We aim to introduce $> 50\%$ of the input power into the cavity with the Gaussian converter
and the focusing lens by matching beam waist position, beam size, beam position,
and the transverse beam shape of the input power to those of the cavity.
For the best maching, we monitor the input, reflected, and accumulated power 
with pyroelectric power monitors located upstream and downstream of the cavity.
Final optimization is now ongoing.

\section{Source and Detectors}
\label{sec:detection}

\begin{figure}[tb]
\begin{center}
\begin{minipage}{16pc}
\includegraphics[width=16pc]{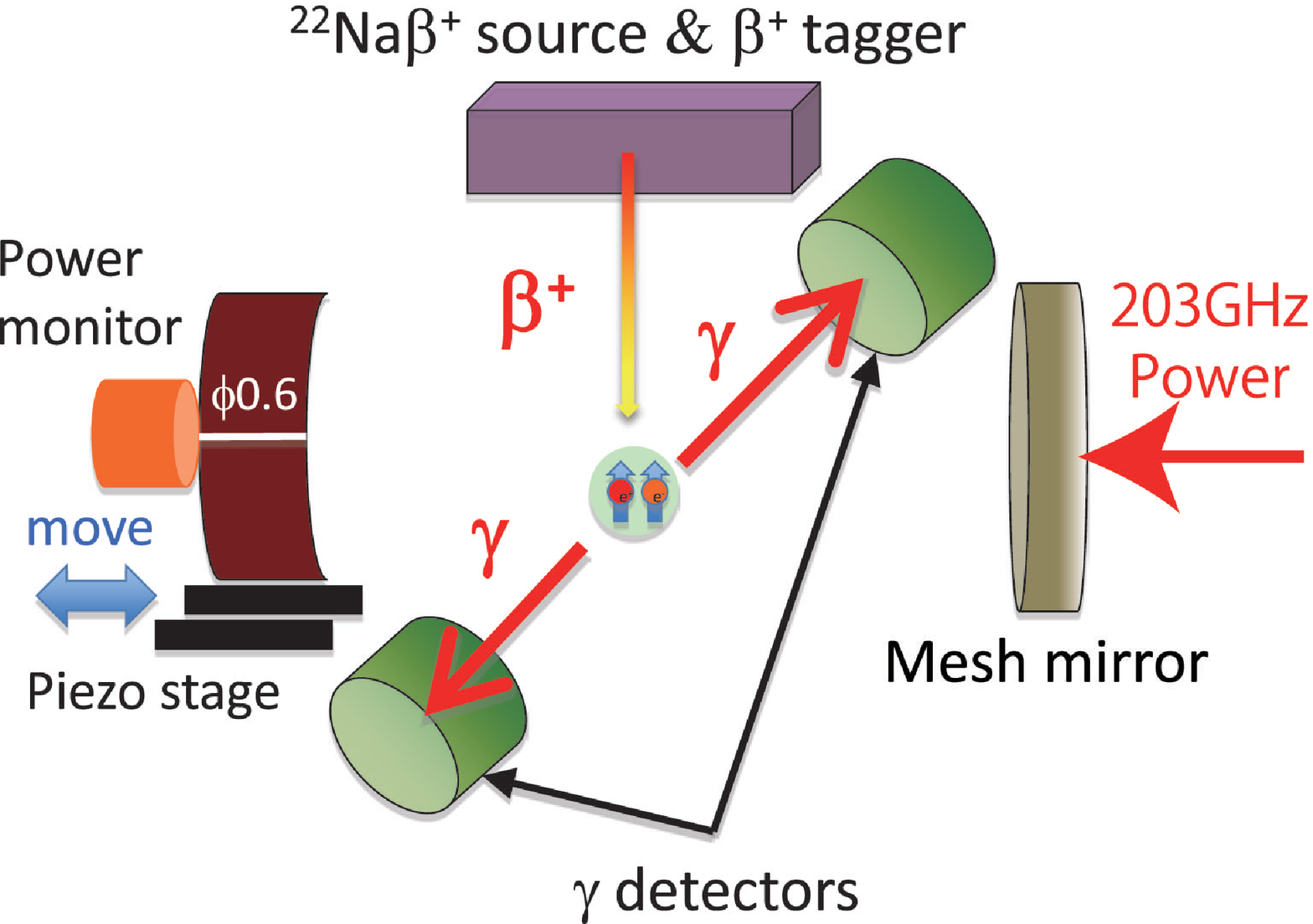}
\end{minipage}\hspace{2pc}%
\begin{minipage}{16pc}
\includegraphics[width=16pc]{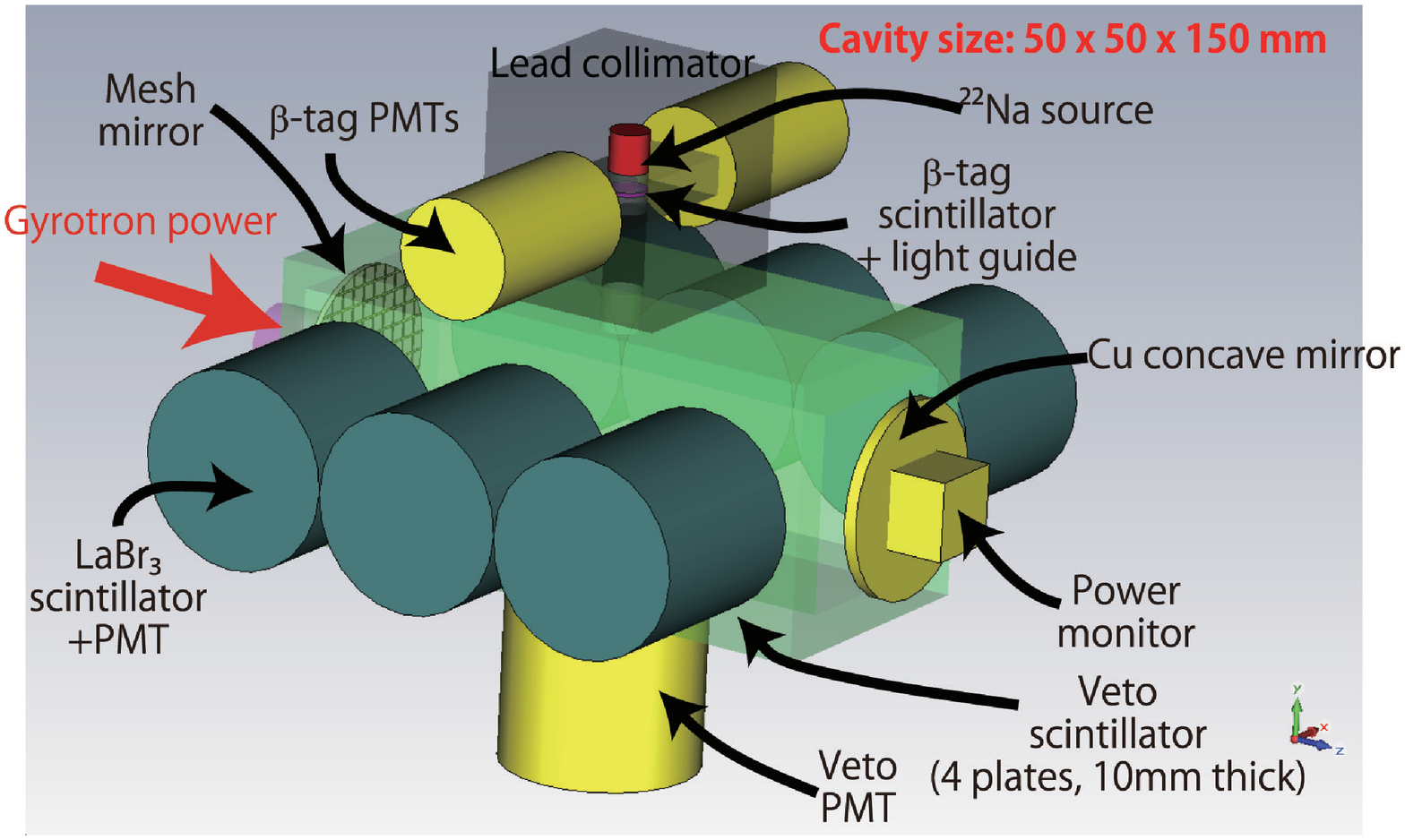}
\end{minipage} 
\caption{\label{fig:detector}Schematic view (left) and 3D drawing (right) of setup geometry around the cavity.}
\end{center}
\end{figure}

Figure \ref{fig:detector} shows the schematic view of the detection system (current plan).
A sodium-22 positron source (1 MBq) is located above the cavity.
The emitted positron passes through a plastic scintillator which generates a start trigger
for timing measurement.
The positron then passes through a lead collimator (shielding of prompt background) to reach the cavity.
The cavity is filled with isobutan gas which decelerates the positron by multiple scattering.
Some of the decelerated positrons form a positronium with an electron in the gas molecule.
p-Ps (25\% of all Ps) immediately decays to two 511 keV monochromatic photons as well as positron without forming positronium,
and o-Ps (75\%) decays to three $< 511$ keV photons with longer lifetime of $\sim 140$ ns.
Under the 203 GHz radiation, some of the o-Ps changes into p-Ps by HFS transition.

For detection of the photons from decayed positronium, six LaBr$_3$ crystal scintillator surround the cavity.
The LaBr$_3$ scintillators have very good energy resolution of $\sim 4\%$ which can efficiently 
separate 511 keV photons (evidence of HFS transition) from photons from o-Ps decay,
and also have good timing resolution of $\sim 300$ psec to separate prompt events (annihilation).
Details of the optimization of source and detector geometry are described in \cite{posmol}.

\section{Summary and Plan}
We are developing a Ps-HFS direct observation experiment.
Using a 203 GHz gyrotron with Gaussian mode converter, quasi-TEM00 high power radiation
is efficiently supplied to a Fabry-P\'erot cavity which achieves $> 99\%$ round-trip reflection.
Efficiency of power introduction to the cavity is now being optimized.
After that, we plan to begin our measurement, which should lead to the
first direct HFS transition observation in this summer.


\end{document}